\documentclass[12pt]{article}
\usepackage{amssymb,amsmath}
\usepackage{graphicx}
\usepackage{ulem}
\usepackage{cite}

\begin{document} 
\begin{center} 
{\bf\Large { Extremely large matter radii in $^{49-51}$Ca isotopes and the $0^{+}$ breathing 
mode states  of $^{48}$Ca}} 
\end{center}
\vskip .1 cm 
\begin{center}
{\bf Syed Afsar Abbas}\\ 
Centre for Theoretical Physics, Jamia Millia Islamia University, New Delhi-110025, India\\
(email: drafsarabbas@gmail.com)\\
\vskip .1 cm
{\bf Anisul Ain Usmani, Usuf Rahaman}\\
Department of Physics, Aligarh Muslim University, Aligarh-202002, India
\end{center} 

\vspace{.05in} 
\begin{center} 
{\bf Abstract}
\end{center}
Through inelastic scattering cross section measurements for $^{42-51}$Ca on
a carbon target at 280 MeV/nucleon recently, Tanaka {\it et al.} [ arXiv:1911.05262 [nucl-ex]], 
in a very significant experiment, have demonstrated
large swelling of doubly magic $^{48}$Ca core in calcium isotopes beyond N=28. 
The matter radii observed in these experiments, are surprisingly much larger than the corresponding, already amazingly large 
charge radii of the same
calcium isotopes, by Garcia {\it et al.}
[Nat. Phys. 12 (2016) 594]. Here we propose a novel solution, wherein the breathing mode
states $0^{+}$ of $^{48}$Ca, provide a global and consistent solution of this matter radii conundrum.

\vskip .8 cm
\vspace{0.1in} 

\vspace{0.2in}

\newpage

Recently, Tanaka {\it et al.} ~\cite{Tanaka2019} have performed measurements of interaction 
cross section, $\sigma_{I}$  for $^{42-51}$Ca, on a neutral carbon target at 280 MeV/nucleon, 
in a groundgreaking first time experiment. Interaction cross section provides informations regarding RMS radius
of nucleon density distributions $<r^{2}_{m}>^{\frac{1}{2}}$, which is referred to as the
``matter radius'' here. This experiment complements the recent determination of charge
radii of the same calcium isotopes (plus $^{52}$Ca also), recently done by 
Garcia {\it et al.}~\cite{Garcia2016}. Surprisingly large charge radii were observed
in calcium isotopes $^{49}$Ca to $^{52}$Ca by Garcia {\it et al.}~\cite{Garcia2016}.
That has presented a great challenge to theoretical physics. Already a conundrum
as it was, in addition, Tanaka {\it et al.} ~\cite{Tanaka2019} have found that the matter radii
of the corresponding isotopes $^{49}$Ca to $^{51}$Ca, are amazingly, 
even much larger than the
corresponding charge radii. This presents a most challenging problem to nuclear 
structure physics.

Through meticulous calculations, Tanaka {\it et al.}~\cite{Tanaka2019} have 
demonstrated that these very large matter radii, cannot be taken as being due to one-,
two- and three-neutron haloes. They have also shown that this cannot be considered as
a deformation effect on these nuclei. What they have convincingly demonstrated
however is, that this is due to substantial swelling of the bare $^{48}$Ca 
core in these calcium isotopes.

So what can be the cause of this swelling of $^{48}$Ca core in these isotopes?
Getting motivation from the concept of E2 effective charge in nucleii, here
we propose a model wherein the breathing mode states $0^{+}$ provide a 
consistent understanding of this puzzling situation made explicit by 
Tanaka {\it et al.} ~\cite{Tanaka2019}.
One may ask, will this large enhancement in matter radii 
continue unabated as we add more and more neutrons to 
$^{48}$Ca core? And whether this will present us with a new calcium-radii-catastrophe?
Our model makes prediction that experimentalists shall find continuous large 
swelling of radii in $^{52}$Ca, $^{53}$Ca and $^{54}$Ca nuclei. 
After which there shall be 
no swelling. Hence, we shall show that our model is not only able to provide
a microscopic  understanding of what has already been demonstrated by 
Tanaka {\it et al.}, but provide new challenges for experimentalists 
to confirm in the future.

The idea of effective charge stems from a desire to explain electromagnetic 
transitions and moments totally within the framework of the shell model.
For example,  we renormalize the quadrupole operator
$\sum_{p}r^{2}Y_{2,0}$ to $(1+e_{p})\sum_{p}r^{2}Y_{2,0} + e_{n}\sum_{n}r^{2}Y_{2,0}$
where the effective charge correction for proton and neutron are 
$e_{p}$ and $e_{n}$ respectively. Note that the total effective charge of proton is $(1+e_{p})$.
It has been emperically determined that $e_{n} \approx +\frac{1}{2}$ and 
$e_{p} \approx +\frac{1}{2}$~\cite{Shalit1959,Abbas1980,Zamick1977}. One has to visualize, for
example, that in $^{17}$O, the $9^{th}$ valence neutron polarizes the $^{16}$O
core so strongly, that it gets highly deformed. Then we transfer this large deformation,
phenomenologicaly by treating the uncharged valence neutron 
to become a highly charged entity of magnitude 
$+\frac{1}{2}$, while still treating the core $^{16}$O as being a ground state spherical
nucleus~\cite{Shalit1959,Abbas1980,Zamick1977}.

In a similar manner, we treat here the $29^{th}$ valence neutron in 
$^{49}_{20}{\rm Ca}_{29}$, to sit outside the magical core nucleus
$^{48}_{20}{\rm Ca}_{28}$ and have an analogous strong deforming effect on it.
This "deforming'' effect, as ascertained by Tanaka {\it et al.}~\cite{Tanaka2019},
is not like that of  $^{16}$O in the case of effective charge  of $^{17}$O, but to 
``polarize'' it so that the core nucleus $^{48}_{20}{\rm Ca}_{28}$ 
``breathes'' out from the ground state 
$0^{+}$, to the first excited $0^{+}_{1}$ collective state of 
$^{48}_{20}{\rm Ca}_{28}$. As a consequence we treat the 
$29^{th}$ valence neutron to just sit smugly on the surface,
while the core has expanded out substantially to mock up the $0^{+}_{1}$ collective
state. This is its ground state now. Thus this is the source of large matter radius as observed 
by Tanaka {\it et al.} in $^{49}$Ca.

Next, in $^{50}_{20}{\rm Ca}_{30}$, now two neutrons sit smugly on the surface of the
core $^{48}_{20}{\rm Ca}_{28}$ nucleus, which then gets even more strongly ``polarized'' by these
(two neutrons), which then breathes out to the next $0^{+}_{2}$ collective state in 
$^{48}_{20}{\rm Ca}_{28}$, thereby providing a much larger matter radius. Similarly in 
$^{51}_{20}{\rm Ca}_{31}$, the valence neutrons sitting smugly on the surface of 
$^{48}_{20}{\rm Ca}_{28}$ would have induced still larger polarization of the core, to breath
out to the next collective $0^{+}_{3}$ state of $^{48}$Ca.

So having explained very large matter radii of $^{49}$Ca, $^{50}$Ca, $^{51}$Ca, we ask as to how long would this large expansion continue? That is,
will this expansion continue in uncontrolled manner in 
$^{52}$Ca, $^{53}$Ca, $^{54}$Ca,..., $^{59}$Ca, $^{60}$Ca? If so, we would have at hand, a 
genuine calcium-radii-catastrophe! One may be reminded of, the end $19^{th}$ century Ultraviolet Catastrophe of the black-body radiation.

As per the phenomenological model proposed here, the valence neutrons on top of N=28 in 
$^{48}_{20}{\rm Ca}_{28}$ would continue to polarize this core strongly, 
so as to go to higher and higher breathing mode states of
$^{48}_{20}{\rm Ca}_{28}$, as long as these states are available. As per data 
displayed below in Table 1, we give the $0^{+}_{n}$ excited breathing mode states of
$^{48}_{20}{\rm Ca}_{28}$, which have been determined experimentally
~\cite{Fujita1988,Burrows1993,nndc}.

\begin{table}
\centering
\caption{Experimental breathing mode states $0^{+}$ of $^{48}$Ca}
\vspace{.1in}
\renewcommand{\tabcolsep}{1.0cm}
\renewcommand{\arraystretch}{1.5}

\begin{tabular}{cc}
\hline\hline
$0^{+}_{1}$ & 4.284 MeV \\
$0^{+}_{2}$ & 5.461 MeV \\
$0^{+}_{3}$ & 11.945 MeV \\
$0^{+'}_{3}$ & 11.967 MeV \\
$0^{+'}_{4}$ & 12.318 MeV \\
$0^{+'}_{5}$ & 12.565 MeV \\
$0^{+'}_{6}$ & 12.869 MeV \\
\hline\hline
\label{tab1}
\end{tabular}
\end{table}

The breathing mode states $0^{+}_{n}$ of $^{48}$Ca have been taken 
from Ref.~\cite{Fujita1988,nndc}. Note the $0^{+}_{3}$ and $0^{+'}_{3}$ at 
11.945 MeV and 11.967 MeV respectively, appear as almost being degenerate. 
The role of these two as being one, is made 
sharper by the fact that in Burrows~\cite{Burrows1993},
$\sigma(\theta)$ in $(p,p^{'})$, $(\alpha,\alpha^{'})$ show oscillatory
pattern and are well fitted by DWBA by assuming and using $0^{+'}_{3}$, $0^{+'}_{4}$, 
$0^{+'}_{5}$, $0^{+'}_{6}$ states (while $0^{+}_{3}$ is not used). 
Note that it is possible that Giant Quadrupole
Resonance states and Giant Monopole Resonance states (breathing mode states) may mingle
with each other in a wide range of excitation energy in $^{48}$Ca ~\cite{Fujita1988}.
Thus if $0^{+}_{3}$ and $0^{+'}_{3}$ states are treated as one, 
the total number of breathing mode states
in $^{48}$Ca are six.

Thus the three observed large matter radii as obtained by Tanaka {\it et al.}~\cite{Tanaka2019}
in $^{49}$Ca, $^{50}$Ca, $^{51}$Ca, would be explained by the three breathing mode states
in $^{48}$Ca at $0^{+}_{1}$ = 4.284 MeV, $0^{+}_{2}$ = 5.461 MeV, 
$0^{+}_{3}$ = 11.945/11.967 MeV. Given the fact that we have three more states
$0^{+}_{4}$ = 12.318 MeV, $0^{+}_{5}$ = 12.565 MeV, $0^{+}_{6}$ = 12.869 MeV, our model
here predicts that one would observe still larger matter radii in 
$^{52}$Ca, $^{53}$Ca, $^{54}$Ca. Thus this expansion will stop at N=34, and
after that there shall be no further swelling, as per our phenomenological model prediction here.


\vskip .5 cm
\begin{center}
\end{center}

\end{document}